\documentclass[%
 reprint,
nofootinbib,
 amsmath,amssymb,
onecolumn
]{revtex4-2}

\usepackage{graphicx}
\usepackage{bm}

\usepackage[
total={6.5in,9.75in}, top=0.9 in, left=0.9in, includefoot,
]{geometry}
\usepackage{color}

\usepackage[normalem]{ulem}

\usepackage{tikz,xcolor,hyperref}
\hypersetup{
    colorlinks,
    linkcolor={black!80!black},
    citecolor={black!50!black},
    urlcolor={black!80!black}
}

\definecolor{lime}{HTML}{A6CE39}
\DeclareRobustCommand{\orcidicon}{
	\begin{tikzpicture}
	\draw[lime, fill=lime] (0,0) 
	circle [radius=0.16] 
	node[white] {{\fontfamily{qag}\selectfont \tiny ID}};
	\draw[white, fill=white] (-0.0625,0.095) 
	circle [radius=0.007];
	\end{tikzpicture}
	\hspace{-2mm}
}
	
\foreach \x in {A, ..., Z}{\expandafter\xdef\csname orcid\x\endcsname{\noexpand\href{https://orcid.org/\csname orcidauthor\x\endcsname}
			{\noexpand\orcidicon}}
}

\begin{document}

\preprint{APS/123-QED}



\title{Waves dictate the yo-yoing decay of a viscoelastic mixing layer}


\author{Giulio Foggi Rota\orcidB{}}
\thanks{These authors contributed equally to this work.}
\affiliation{Complex Fluids and Flows Unit, Okinawa Institute of Science and Technology Graduate University, Okinawa 904-0495, Japan}
\author{Piyush Garg\orcidC{}}
\thanks{These authors contributed equally to this work.}
\affiliation{Complex Fluids and Flows Unit, Okinawa Institute of Science and Technology Graduate University, Okinawa 904-0495, Japan}
\author{Jason Tang\orcidA{}}
\affiliation{Complex Fluids and Flows Unit, Okinawa Institute of Science and Technology Graduate University, Okinawa 904-0495, Japan}
\author{Marco Edoardo Rosti\orcidD{}}%
\email{marco.rosti@oist.jp}
\affiliation{Complex Fluids and Flows Unit, Okinawa Institute of Science and Technology Graduate University, Okinawa 904-0495, Japan}

\begin{abstract}

We find that waves develop in a time-decaying mixing layer of viscoelastic fluid, leading the mean--flow to yo-yo. This is in sharp contrast with Newtonian fluids, where laminar mixing layers evolve monotonically. We combine direct numerical simulations with a theoretical analysis of the energy budget for the flow to uncover the underlying physical mechanism. The yo-yoing of the mean--flow is shown to be driven by the elastic {polymers} injecting energy into the fluid and, in turn, being rotated by the large-scale mean shear. We then provide the mathematical model of the problem and solve it analytically, finding wave solutions with non--linear dispersion predicting the period of the yo-yoing and the parameter range where it occurs. 
As decaying mixing layers are one of the simplest and canonical examples of unsteady flows, the phenomenon identified here explains the anomalies recently observed in  experiments of unsteady viscoelastic flows in complex geometries.

\end{abstract}
\maketitle

\section{Introduction}

Most fluids, whether in biological, industrial or geophysical settings, are not simply Newtonian but comprise of an internal microstructure \cite{bird1987dynamics, balmforth2001geophysical, morozov2015introduction}. Understanding the flows of such `complex' fluids is still a challenging task. Viscoelastic fluids, such as polymer solutions and melts, have drawn a large amount of interest as a prototypical and widespread example \cite{bird1987dynamics, larson2013constitutive, datta2022perspectives, steinberg2021elastic}. They respond to external forcing with an elastic response, in addition to the usual viscous Newtonian stress, and  as a result exhibit a variety of surprising dynamics - from novel forms of pattern formation and macro-scale behaviour \cite{shaqfeh1996purely, bergeron2000controlling, datta2022perspectives} to wave propagation \cite{varshney2019elastic, de2023canopy} and even inertialess, elasticity-driven, turbulence \cite{groisman2000elastic, steinberg2021elastic, singh2024intermittency}.  

The {polymers relax to their equilibrium state} on an intrinsic time scale ($\tau$), while the fluid and {the polymers} are driven out of equilibrium by the imposed shear rate ($\dot{\gamma}$). At large Deborah numbers ($De = \dot{\gamma} \tau $), the {polymers} strongly affect the flow, which then further perturbs {them}, leading to a strong coupling between the two. This two-way interaction has meant that even canonical shearing flows of viscoelastic fluids at large $De$, such as pressure-driven flows and mixing layers, are not yet well understood. Progress has concentrated on pressure-driven flows, where transition to a spatio--temporally chaotic state is observed at large $De$ \cite{pan2013nonlinear, lellep2024purely, rota2024unified}. This `elastic turbulence' state is sustained by a balance between elasticity and viscous dissipation, without any contribution from fluid inertia \cite{steinberg2021elastic, singh2024intermittency}. The transition to elastic turbulence occurs due to {novel elasticity--driven instabilities and associated coherent structures \cite{boffetta-etal-2005, garg2018viscoelastic, page2020exact, morozov2022coherent, lewy-kerswell-2025-1}}. 

Surprisingly, even less is known about viscoelastic mixing layers, characterised by momentum transfer between co-moving streams with differing velocities. Since a mixing layer describes pattern formation in a time dependent mean--flow, understanding its roll-up has been foundational to the dynamics of Newtonian fluids in a wide variety of settings \cite{ho1984perturbed, peltier2003mixing}. For viscoelastic fluids, early work on mixing layers focused on the inertial regime and concluded that viscoelasticity primarily stabilises the inertia driven roll--up \cite{azaiez1994linear, azaiez1994numerical, ray2014absolute}. Recent experiments, in two different geometries, have examined the large $De$ mixing layer - one set for a mixing layer generated behind an obstacle inside a channel \cite{varshney2018mixing, jha2021elastically} and the second set for the flow over a rigid microcanopy \cite{de2023canopy, lopez2025canopy}. Both sets of experiments reported anomalous vorticity generation and turbulent--like velocity fluctuations. However, no theoretical or numerical studies exist so far in the parameter regime corresponding to these experiments, and an understanding of the underlying physical mechanism for the observations is still lacking. Furthermore, in existing studies, it is widely assumed that the time evolution of the large-scale mean--flow of the mixing layer is not altered by viscoelasticity \cite{azaiez1994linear, azaiez1994numerical, yu2004three, ray2014absolute}.  

In contrast, here we show that a viscoelastic mixing layer does not always evolve monotonically, but instead it can yo-yo. The coupling of the {polymers} to the mean--flow leads to multiple reversals of the mixing layer. We combine direct numerical simulations with a theoretical analysis of the energy budget to uncover the underlying physical dynamics. We also reproduce the phenomenon using a reduced 1D-model for the streamwise averaged fields, finding analytical wave-like solutions with non--linear dispersion. Noticeably, the model accurately predicts the period of the yo-yoing and the parameter range where it occurs. We thus show that viscoelasticity can completely alter the time evolution of the large-scales of a mixing layer, sustaining wave--dominated dynamics. Finally, we further discuss the relevance of the described mechanism to recent experimental observations \cite{varshney2018mixing, jha2021elastically, de2023canopy, lopez2025canopy}. 

\section{Results}
\subsection{Problem statement}

For a viscoelastic fluid, the velocity field $\boldsymbol{u}$ is governed by the Navier-Stokes equation along with the continuity equation for incompressibility. The {polymers} are assumed to deform affinely with the flow and hence are modelled using the {Oldroyd-B equation \cite{bird1987dynamics, hinch2021oldroyd, hillebrand-etal-2025, foggirota-etal-2026}} through the evolution of their conformation tensor $\boldsymbol{C}$. The governing equations are thus
\begin{equation}
\rho \left( \partial_t \boldsymbol{u} +  \left( \boldsymbol{u} \cdot \boldsymbol{\nabla} \right) \boldsymbol{u} \right)=-\boldsymbol{\nabla} p+ \mu_f \nabla^2 {\boldsymbol{u}} +\frac{\mu_p}{\tau} \boldsymbol{\nabla} \cdot \boldsymbol{C},
\end{equation}
\begin{equation}
\boldsymbol{\nabla} \cdot \boldsymbol{u} =0,
\end{equation}
\begin{equation}
\partial_t \boldsymbol{C} + \boldsymbol{u} \cdot \boldsymbol{\nabla} \boldsymbol{C} = \boldsymbol{C} \cdot \boldsymbol{\nabla}  \boldsymbol{u} + \boldsymbol{\nabla} \boldsymbol{u}^{T} \cdot \boldsymbol{C} - \frac{\boldsymbol{C}-\textbf{I}}{\tau},
\end{equation}
where $\rho$ is the fluid density, $\mu_f$ the fluid viscosity, $\mu_p$ the additional {polymeric} viscosity, and $\tau$ its relaxation time. We solve these equations in two dimensions where $x$ and $y$ are taken to be the streamwise and gradient directions respectively. See the methods section for further details on the numerics. {In the following,} we write the average of any variable $f$ over the whole domain as $\langle f\rangle$, while $\langle f\rangle_x$ corresponds just the streamwise average.  {Streamwise averaged values at the centreline are denoted with $\langle f\rangle_c$.}

\begin{figure}
	\includegraphics[width=.475\textwidth]{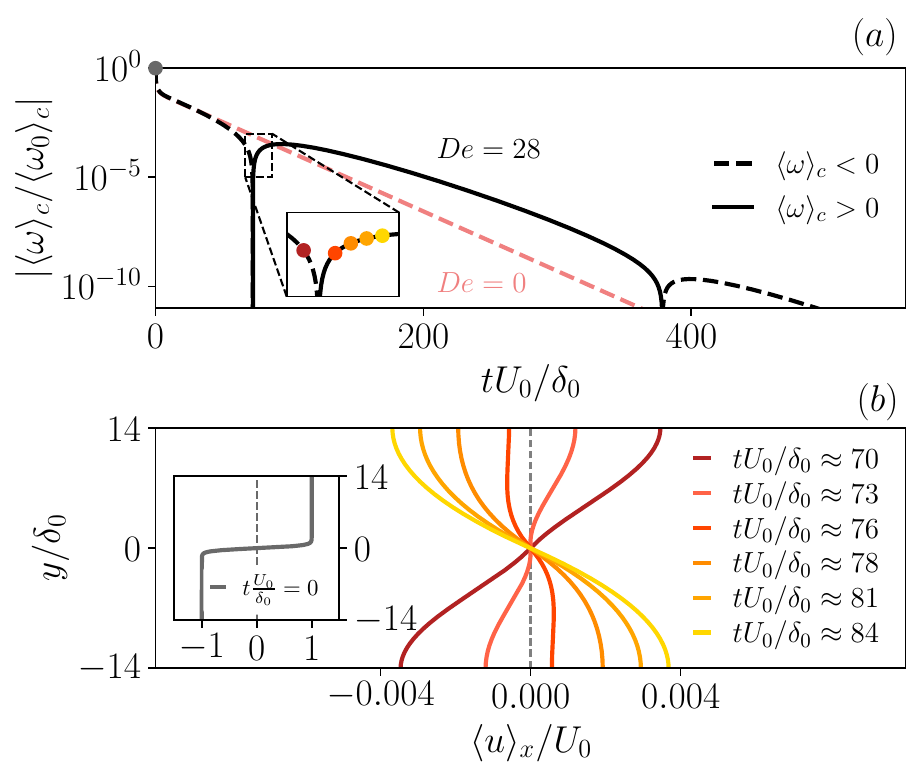}
	\includegraphics[width=.475\textwidth]{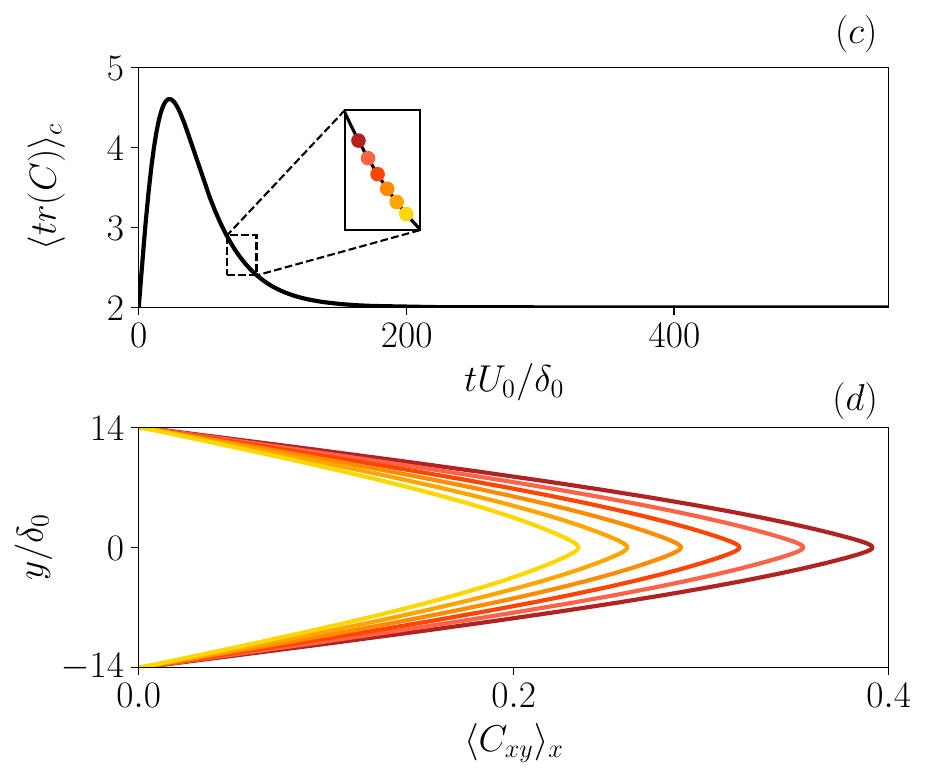}
	\includegraphics[width=.93\textwidth]{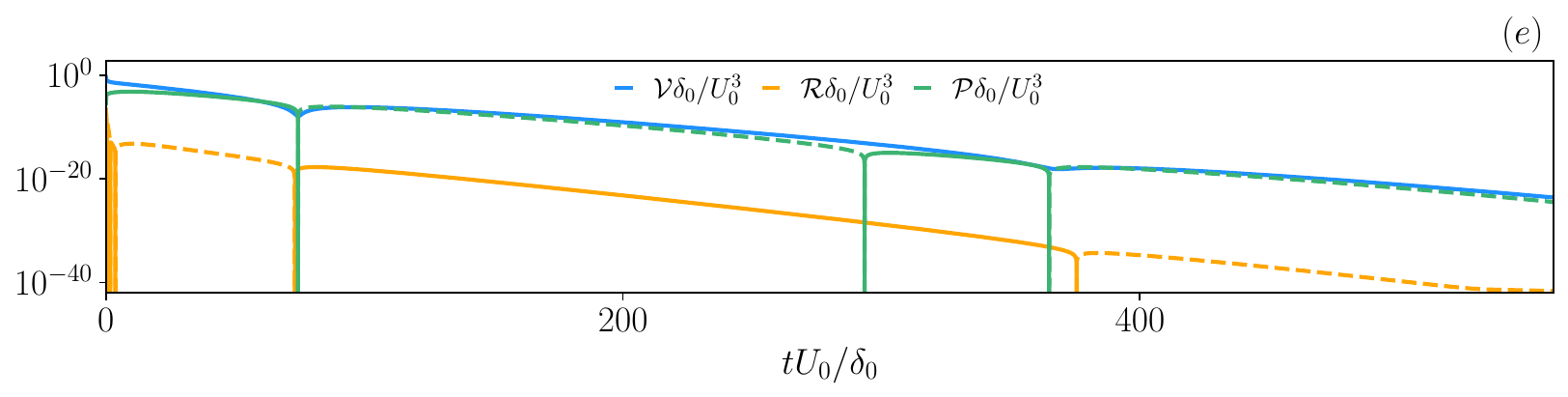}
	\caption{\textbf{Yo-yoing of the viscoelastic mixing layer.} $(a)$ Evolution of the absolute value of the {mean centreline vorticity $| \langle \omega \rangle_c  / \langle \omega_0 \rangle_c |$} with the non-dimensional time $tU_{0}/\delta_0$ for the viscoelastic ($De = 28$) and Newtonian cases ($De = 0$). The inset is zoomed--in near the overturning time $t U_0/\delta_0 \sim 74.6$, where {$\langle \omega \rangle_c $} changes sign. $(b)$ The mean streamwise velocity field $\langle u \rangle_x/U_0$ versus the gradient direction $y/\delta_0$ at various times $tU_{0}/\delta_0$ near $74.6$; the inset shows the mixing layer at time $t=0$. $(c)$ Evolution of the {centreline} trace of the {conformation tensor} {$\langle tr(C) \rangle_c$} over time $tU_0/\delta_0$. The inset is zoomed--in near the overturning time $t U_{0}/\delta_0 \sim 74.6$. $(d)$ The spatial variation of the streamwise-averaged $\langle C_{xy} \rangle_x$ versus $y/\delta_0$ at various times $tU_0/\delta_0$ near $74.6$. $(e)$ Evolution of the spatially averaged viscous dissipation $(\mathcal{V}\delta_0/U_0^3)$, turbulent production $(\mathcal{R}\delta_0/U_0^3)$ and microstructure contributions $(\mathcal{P}\delta_0/U_0^3)$ in the energy balance of the mean kinetic energy of the fluid over time $tU_0/\delta_0$. Continuous lines denote positive values, dashed lines negative values.}
    \label{fig:1}
\end{figure}

We initialise the flow with a mixing layer of thickness $2\delta_0$ confined to the centre of {a squared domain with edge $L=28\delta_0$}, and superimpose a harmonic perturbation $\boldsymbol{ u}_t$ with small amplitude and zero mean, so that at $t = 0$, $ \boldsymbol{u} =  U_0 \tanh \left( 2 y/\delta_0 \right) \boldsymbol{\hat{x}} + \boldsymbol{u}_t$ and $\langle \boldsymbol{ u}_t \rangle_x = 0$ (shown in the inset of figure \ref{fig:1}$(b)$). Periodicity is enforced along the streamwise direction, while free-slip is imposed at the top and bottom boundaries, where we also impose $\partial ^2 \boldsymbol{C}/\partial y^2=\boldsymbol{0}$. The flow is parametrised by the Reynolds number $Re = \rho U_0 \delta_0/(\mu_f +\mu_p))$, the Deborah number $De = \tau U_0/\delta_0 $ and the viscosity ratio $\beta = \mu_f/(\mu_f+ \mu_p)$, where $U_0$ is the initial velocity scale. For most of the results presented, we choose a parameter set of $Re = 0.2$, $De = 28$ and $\beta = 0.9$, where inertial effects are expected to be negligible. We choose the length scale $\delta_0$ and time scale $\delta_0/U_0$ to present the {simulation} results. 

\subsection{Physics of the yo-yoing}
Figure \ref{fig:1}$(a)$ shows the {absolute} variation of the (normalised) mean centreline vorticity {$|\langle \omega \rangle_c / \langle \omega_0\rangle_c |$} over time $tU_0/\delta_0$, for both the Newtonian ($De = 0$) and viscoelastic ($De = 28$) cases,  where {$\langle \omega_0 \rangle_c$} is the value of the mean {centreline} vorticity at $t=0$. {After a sharper initial adaptation,} the vorticity decays as the mixing layer expands due to momentum diffusion, similarly to Newtonian fluids. However, we observe a striking departure from this behaviour as the mean {centreline} vorticity {drops}, changes sign, and then \textit{increases} around $tU_0/\delta_0 \sim 74.6$. The mean {centreline} vorticity has thus effectively flipped sign in a very short duration {(and so does its volume averaged value)} - implying that the mixing layer has entirely overturned with the streams now flowing in opposite directions to how they started. After changing sign, the mean {centreline} vorticity resumes a similar time decay as before the overturning. {We confirmed that the exponential decay rate of the mean centreline vorticity matches that of the velocity, being half that of the kinetic energy in the system. The evolution of relevant flow quantities can thus be inferred from it throughout the decay.}
Figure \ref{fig:1}$(b)$ shows the gradient direction variation of the streamwise--averaged velocity field, $\langle u \rangle_x$, for times just before and after the overturning; the velocity streams indeed reverse direction. We eventually see another overturning of the mixing layer around $tU_0/\delta_0= 365.4$, even as overall the vorticity continues to decay. The velocity field of a viscoelastic mixing layer thus yo-yos over time even as the momentum diffuses across it. This is in sharp contrast to Newtonian mixing layers, where the mean {centreline} vorticity monotonically decays over time. 

Next, we look at how the {polymer conformation} evolves. Figure \ref{fig:1}$(c)$ shows the time evolution of {$\langle tr(C) \rangle_c = \langle C_{xx} + C_{yy} \rangle_c$}, which is a measure of the {polymer stretching} {at the centreline}. We start with the {polymers} at equilibrium, so {$\langle tr(C) \rangle_c = 2$} at $t= 0$. Initially, the {polymers are} stretched by the flow and absorb energy from it. Eventually, {$\langle tr(C) \rangle_c$} starts to decrease, implying that the {polymers are} slowly relaxing back to their equilibrium state. Since the mean--flow is approximately unidirectional, only $\langle C_{xx} \rangle_x$ and $\langle C_{xy} \rangle_x$ have a non--trivial spatial variation over time, while $\langle C_{yy} \rangle_x \approx 1$ for all times; $\langle C_{xy} \rangle_x$ is a measure of the orientation of the {polymers}. Figure \ref{fig:1}$(d)$ shows the gradient direction $(y/\delta_0)$ variation of the shear component of the stress due to the {polymers}, $\langle C_{xy}\rangle_x$, for {the same time instants $tU_0/\delta_0$ in figure \ref{fig:1}$(b)$} near the overturning time $tU_0/\delta_0 \sim 74.6$. {While the velocity profiles reverse around that time, $\langle C_{xy} \rangle_x$ does not, and undergoes a monotonic decay with its maximum always located at the centreline} $(y = 0)$, where the mixing layer initially had the strongest velocity gradients. {As reported in the Supplementary Information, $\langle C_{xx} \rangle_x$, exhibits a similar behaviour and non-monotonic modulation of the polymer conformation instead occurs at a different time from that of flow reversal.}

Since there is no external source of energy in the mixing layer, the yo-yoing of the mean velocity field has to be driven by the transfer of energy to/from the {polymers}. The equation governing the time evolution of the mean kinetic energy ($E = \langle \langle \boldsymbol{u} \rangle_x \cdot \langle \boldsymbol{u} \rangle_x \rangle/2$) can be derived from the governing equations as,
\begin{equation}
  \frac{\mathrm{d}E}{\mathrm{d}t} = -  \biggl \langle \left( 2 \frac{\mu_f}{\rho} \langle \mathbf{S} \rangle_x  -   \langle {\mathbf{u}_t \mathbf{u}_t} \rangle_x + \frac{\mu_p}{\rho \tau} \langle \mathbf{C} \rangle_x \right) \boldsymbol{:} \langle \nabla  \mathbf{u} \rangle_x  \biggr\rangle,
  \label{eq:KE}
\end{equation}
where $\boldsymbol{u_t}  = \boldsymbol{u}  - \langle \boldsymbol{u}  \rangle_x$, see the Supplementary Information for a detailed derivation. The first and second term on the RHS are the viscous dissipation ($\mathcal{V}$) and the turbulent production from the Reynolds stresses ($\mathcal{R}$), familiar from Newtonian fluids. The last term in the energy balance equation -- which represents the additional energy transfer mechanism for viscoelastic fluids through the coupling with the {polymers} ($\mathcal{P} = \frac{\mu_p}{\rho \tau} \langle \langle \mathbf{C} \rangle_x  \boldsymbol{:} \langle \nabla  \mathbf{u} \rangle_x \rangle $) -- injects energy into the flow when $\mathcal{P }  < 0$ and dissipates energy when $  \mathcal{P }  > 0$. In the Supplementary Information, we show that the energy balance is exactly satisfied for the time interval considered, thus establishing the validity of the numerical scheme used.

\begin{figure}
        \includegraphics[width=0.75\textwidth]{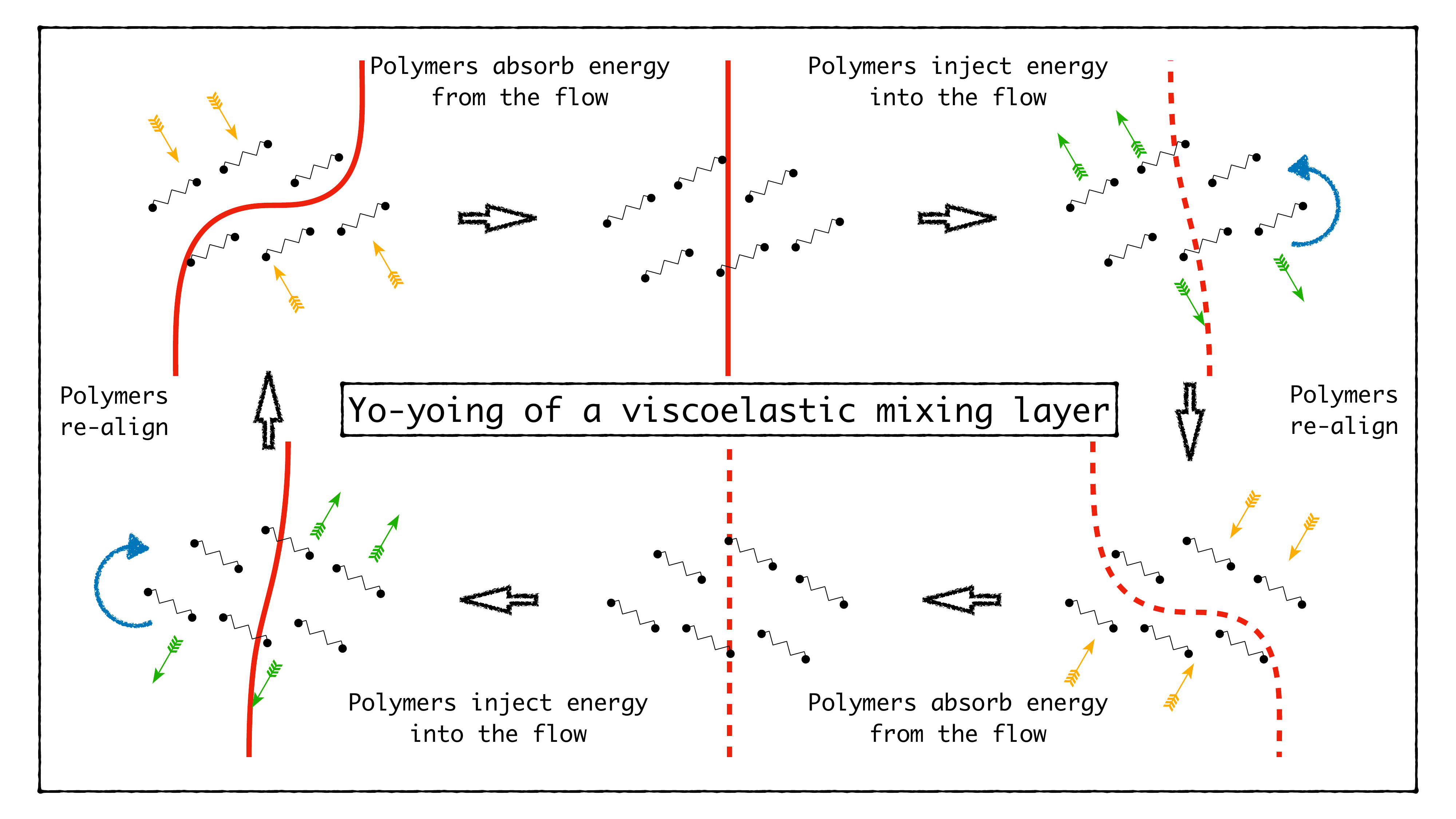}
		\caption{\textbf{Schematic illustrating the physical mechanism.} Yellow and green arrows indicate the direction of energy exchange between the fluid and the {polymers} (yellow: to the {polymers}; green: to the fluid).}
    \label{fig:schm}
\end{figure}

Figure \ref{fig:1}$(e)$ shows the time evolution of the three terms. The turbulent production $\mathcal{R}$ has a negligible contribution, thus confirming that the advective non--linearity is irrelevant; in the Supplementary Information we have further verified this by completely removing it from the evolution equations. The viscous dissipation $\mathcal{V}$ stays positive definite always, as expected. The yo-yoing dynamics is thus entirely driven by the {polymer} coupling term $\mathcal{P}$. The time evolution of $\mathcal{P}$ can help us infer a mechanism for the yo-yoing, which is summarised as a schematic in figure \ref{fig:schm}.  Since the mean--flow is approximately unidirectional in the mixing layer, the {polymer} coupling term $\mathcal{P}$ simplifies to $\mathcal{P } \approx \frac{\mu_p}{\rho \tau} \langle \langle C_{xy} \rangle_x \langle \mathrm{d}_y u \rangle_x \rangle$. Initially, the {polymers absorb} energy from the flow as {they stretch}, so $\mathcal{P} > 0$ (implying both $\langle \mathrm{d}_y u \rangle >0$ and $\langle C_{xy} \rangle >0$). Hence, the {polymers are} oriented along the flow direction during this time, and act as an additional dissipative mechanism. Around $t U_0 /\delta_0 \sim 74.6$ we see that both the {polymer} term $\mathcal{P}$ and the mean {centreline} vorticity {$\langle \omega \rangle_c$} change sign. For $tU_0/\delta_0 $, slightly larger than $74.6$, the {polymer} term $\mathcal{P}$ (now a production term) exceeds the viscous dissipation $\mathcal{V}$, and hence the mean kinetic energy (and the vorticity) can grow. Thus, at $tU_0/\delta_0 \sim 74.6$ the {polymers start} injecting energy into the fluid and inducing an overturning of the mixing layer with the two fluid streams flowing in opposite directions to before. Crucially, the reversed direction of the mean shear implies that the {polymers are} now oriented against the flow gradient, so $\langle \mathrm{d}_y u \rangle <0$ while $\langle C_{xy} \rangle >0$. Over time, the {polymers} once again become aligned along the velocity gradient of the new mixing layer and start to dissipate energy; this happens around $tU_0/\delta_0 \sim 293.5$ in figure \ref{fig:1}$(e)$, when $\mathcal{P}$ changes sign again (so $\langle C_{xy} \rangle <0$ as shown in the Supplementary Information). This brings us back to a mirror of the initial configuration, which then just sets up another overturning of the mixing layer at $tU_0/\delta_0 \sim 365.4$, and the pattern repeats as sketched in figure \ref{fig:schm}. 

The physical mechanism{, although observed here for the specific case of an Oldroyd-B liquid,} should be readily generalisable to {describe the one-dimensional mean flow dynamics of} any complex fluid with orientable microstructure comprising a mixing layer. Indeed, the yo-yoing is an interplay between the decay of the mixing layer and the reorientation of the microstructure by the large scale flow. Since the microstructure reorients in between the overturning of the mean flow, the multiple reversals rely on the microstructure being out of synchronisation with the induced flow, where the two time scales - decay and reorientation - are comparable. If either process occurs on a different time scale, we can no longer expect multiple reversals to occur. 

However, we can still expect such a two-way coupling to persist for sufficiently elastic fluids over a wide range of parameters. Figure \ref{fig:4}$(a)$ and $(b)$ show the time evolution of the mean {centreline} vorticity upon varying the non-dimensional {polymer} relaxation time $De$ and concentration through the viscosity ratio $\beta$, respectively. For low De $ (2.8)$, there is no reversal of the mixing layer, and the time decay of mean {centreline} vorticity is monotonic like the Newtonian case. We find the reversal at larger $De$, both for $28$ and $280$. In addition, at $De =280$ the decay of the mean {centreline} vorticity of the reversed mixing layer is slower and the mixing layer induced after the overturning is sustained for a long{er} time. Upon increasing $\beta$, the contribution of the {polymers} to the total viscosity increases, so $\beta = \mu_f/(\mu_f +\mu_p)$ decreases. The frequency of the reversals increases with increasing $\beta$ as the energy injection by the {polymers} can more easily overwhelm the viscous dissipation (figure \ref{fig:4}$(b)$). The overall decay of the {mean centreline} vorticity is also slower at smaller $\beta$, since the reversed mixing layer is induced by the {polymers themselves}.

\subsection{Viscoelastic waves}
\begin{figure}
        \centering
        \includegraphics[width=\textwidth]{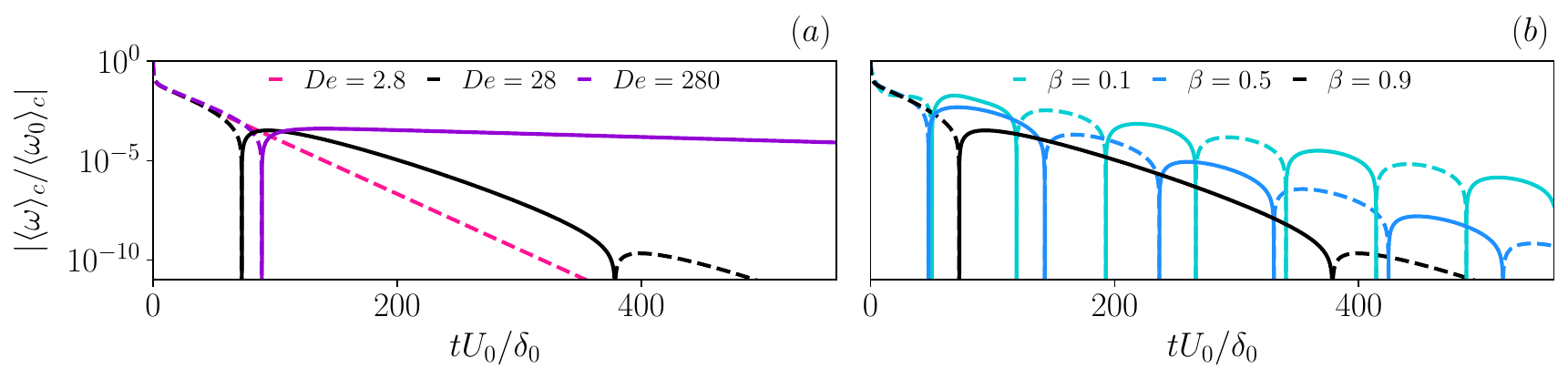}
        \includegraphics[height=.2\textheight]{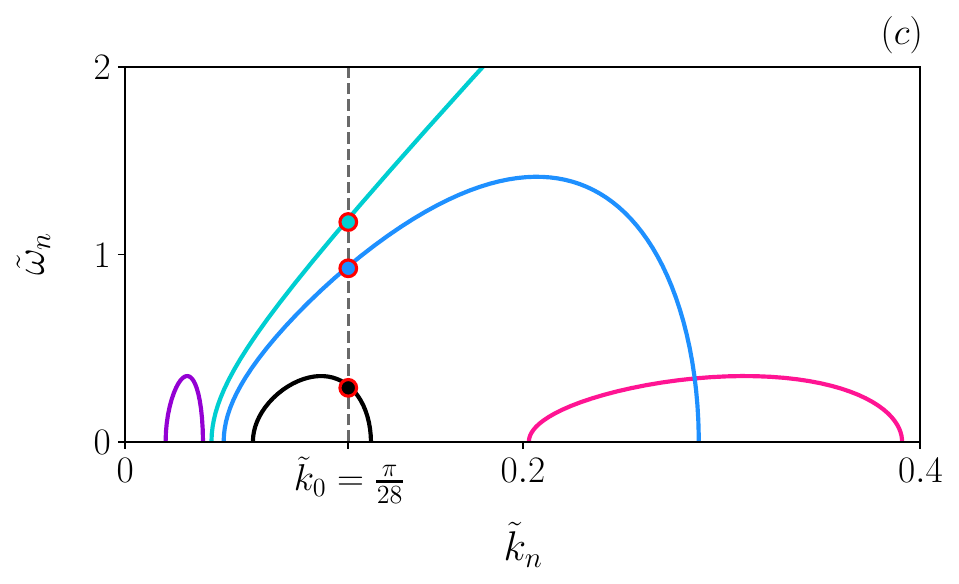}
        \includegraphics[height=.2\textheight]{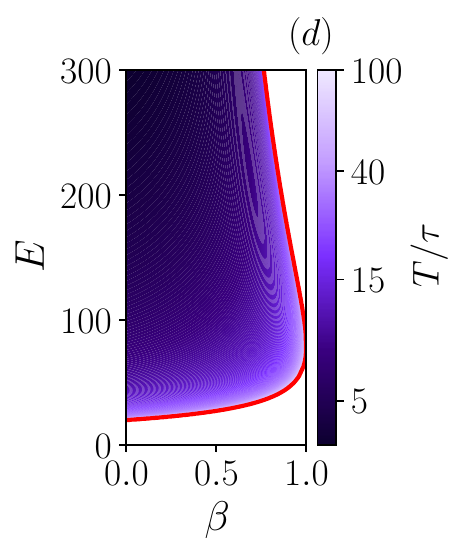}
         \includegraphics[height=.2\textheight]{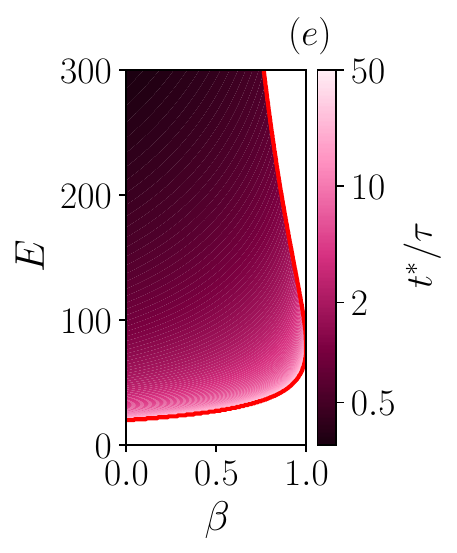}
		\caption{\textbf{Viscoelastic waves.} Evolution of the mean {centreline} vorticity {$|\langle \omega \rangle_c/ \langle\omega_0 \rangle_c |$} over time $tU_0/\delta_0$ for varying $(a)$ Deborah numbers ($De$) with $Re = 0.2$ and $\beta = 0.9$, and $(b)$ viscosity ratios $(\beta)$ with $Re = 0.2$ and $De = 28$. Continuous lines denote positive values, dashed lines negative values. $(c)$ Dispersion relation for the viscoelastic waves periodically overturning the mixing layer as predicted by our simplified flow model, for the parameter choices in $(a,b)$. Overturning at the frequency corresponding to the wavenumber $\tilde{k}_0$ is in remarkable agreement with our simulation results, reported as red--edged dots. The cases at $De=2.8,\enspace 280$ do not yo-yo since their dispersion does not intersect $\tilde{k}_0$. $(d)$ Model prediction for the parameter region where the yo-yoing occurs, i.e. where $\tilde{\omega}_0$ is real, in the same domain of our simulations with extent {$L=28\delta_0$} along the gradient direction. The map is coloured with the period $T$ of the resulting oscillations. { $(e)$ First reversal time estimate under the least decaying wave assumption, i.e. of the first active mode. Note that quantities in the lower panels are scaled as in the model, differently from the reported simulation results on top. Thus, e.g., times differ by multiplication with the $De$ number.}}
    \label{fig:4}
\end{figure}
The streamwise--averaged dynamics so far is seen to be unidirectional, and thus follows a reduced set of equations,
\begin{align}
    \frac{\partial\langle u \rangle_x}{\partial t} &= \frac{\mu_f}{\rho} \frac{\partial^2\langle u \rangle_x}{\partial y^2} +  \frac{\mu_p}{\rho \tau}\frac{\partial \langle C_{xy} \rangle_x}{\partial y}, \nonumber \\ 
    \frac{\partial \langle C_{xy} \rangle_x}{\partial t} &= \langle C_{yy} \rangle_x \frac{\partial\langle u \rangle_x}{\partial y} - \frac{1}{\tau}\langle C_{xy} \rangle_x .
\end{align}
We can further eliminate the {conformation} tensor and write a single equation for the dimensionless velocity $\tilde{u}=\langle u \rangle_x / U_0$ as
\begin{align}
   \frac{\partial^2 \tilde{u} }{\partial \tilde{t}^2} +  \frac{\partial \tilde{u}}{\partial \tilde{t}}   - E  \frac{\partial^2 \tilde{u} }{\partial \tilde{y}^2} = E \beta \frac{\partial^3 \tilde{u} }{\partial \tilde{t} \partial \tilde{y}^2 },
\end{align}
where $\tilde{y}=y/\delta_0$, $\tilde{t}=t /\tau$, $\tilde{u}=u/U_0$, and $E=De/Re = \tau \left( \mu_f + \mu_p \right)/\left(  \rho \delta_0^2 \right) $ is the elasticity number.
Monotonous velocity decay is found for $\beta \in (0,1]$ in the limit $E\rightarrow 0$ / $E\rightarrow +\infty$, for which $\tilde{u}\sim e^{-\tilde{t}}$ / $\tilde{u}\sim e^{-\tilde{t}/\beta}$, respectively (note that, to attain the classical Newtonian decay exponent, the time normalisation needs to be changed). 
Yet, for intermediate values of the elasticity the problem admits oscillatory solutions \cite{zaitsev2002handbook} of the form 
\begin{equation}
   \tilde{u} (\tilde{y},\tilde{t}) = \sum_{n=0}^{+\infty}(a_n e^{\gamma_+(\tilde{k}_n) \tilde{t}} + b_n e^{\gamma_-(\tilde{k}_n) \tilde{t}})B_n(\tilde{y}),
   \label{eq:solution}
\end{equation}
where $B_n$ is a set of sinusoidal base--functions compatible with the boundary conditions, while the coefficients $a_n, b_n$ follow from the initial conditions, projected over the infinite set of discrete modes with wavenumbers {$\tilde{k}_n=(2n+1)\pi/L$} (further details on the derivation are found in the Supplementary Information).  
The temporal dynamics of the mixing layer is governed by the exponents 
\begin{align}
   \gamma_{\pm}(\tilde{k}_n) = -{\alpha} \pm \sqrt{{\alpha^2}-E \tilde{k}_n^2}, \enspace \enspace \enspace {\textrm{with} \enspace  \alpha=\frac{1+\beta E \tilde{k}_n^2 }{2}},
\end{align}
and oscillations emerge when $\gamma_{\pm}$ are complex. 
The yo-yoing mechanism can thus be interpreted as a {linear} wave with frequencies $\tilde{\omega}_n$ obeying the dispersion relation
\begin{equation}
   \tilde{\omega}_n = \sqrt{E \tilde{k}_n^2 - {\alpha^2}},
   \label{eq:dispersion}
\end{equation}
which in the limit of vanishing solvent viscosity ($\beta \rightarrow 0$) reduces to the known dispersion relation for shear waves in solids, $ \tilde{\omega}_n \sim \sqrt{E} \tilde{k}_n$ \cite{achenbach2012wave}. We confirm the relation by predicting the oscillations of the mixing layer in our simulations, as shown in figure \ref{fig:4}$(c)$. 
{By doing so, it is relevant to note that simulation times are scaled with $\delta_0/U_0$ following directly from the initial condition, while frequencies (and periods) from equation~(\ref{eq:dispersion}) are defined in terms of the polymer relaxation time $\tau$. The conversion factor between the two is the $De$ number itself.}
{We assume that the yo-yoing is dominated by the first active mode ($n=0$): the initial velocity profile employed throughout this work, chosen as representative of a realistic shear profile, strongly projects onto the gravest mode, which contains $\sim 84\%$ of the initial kinetic energy. Even though higher modes are initially active, they damp quadratically in their respective wavenumber, so the emerging dynamics is that of mode zero, justifying the \textit{least decaying wave assumption}.}
Model predictions for $\tilde{\omega}_0$ are thus found as intersections between the dispersion relation specialised for the parameter choice and the vertical line at $\tilde{k}_0$. Red--edged dots denote values of $\tilde{\omega}_0$ measured from the simulations in figure \ref{fig:4}$(a,b)$: agreement with the model is outstanding.
{The oscillation frequency (hence the timescale of the overturning) thus appears well captured by relation~(\ref{eq:dispersion}) specialised to the first active mode. ``Unpacking" the elasticity number $E=De/Re=(\tau U_0 / \delta_0)/(\rho U_0 \delta_0/(\mu_f+\mu_p))=\tau (\mu_f+\mu_p) / (\rho \delta_0^2)$, we observe a clear dependence of such frequency from \textit{(i.)} the stress relaxation time of the polymers $\tau$, \textit{(ii.)} the total viscosity $\mu_f+\mu_p$ and density $\rho$, and \textit{(iii.)} the shear layer thickness, i.e., the wavelength of the first active mode (appearing through both $\delta_0$ and $\tilde{k}_0^2$).}
 Next, in figure \ref{fig:4}$(d)$ we fill the region of the parameter space where the first mode exhibits oscillatory dynamics, coloured with the oscillation period $T/\tau=2\pi/\tilde{\omega}_0$. {As expected,} the result is dependent on the vertical extent of the domain through the imposition of the free--slip boundary conditions (further details in the Supplementary Information), and no oscillations are predicted when the domain becomes unbounded. However, these conditions have been used to represent mixing layers in other natural systems, e.g. stratified flows \cite{caulfield2021layering}.
 
{
The time of the first reversal is also a matter of interest. Unfortunately it is not immediate to derive an estimate based on simple arguments, and its computation from the analytical model entails the coupled solution of an infinite set of equations just to attain a prediction dependent on the chosen initial condition. Well aware of its limitations, we thus proceed under the least decaying wave assumption and derive (with simple algebraic manipulations) the first reversal time $t^*$ of the gravest active mode. We thus truncate equation~(\ref{eq:solution}) at $n=0$, compute the vorticity as its spatial gradient, and set to zero its centreline value. This leads to writing $a_0 e^{\gamma_+(\tilde{k}_0)\tilde{t}^*} + b_0 e^{\gamma_-(\tilde{k}_0)\tilde{t}^*}=0$, where {the coefficients $a_0$ and $b_0$ carry information on the amplitude and phase of the shear layer evolution at the gravest wavenumber. They emerge as a combination of the normalised projections $c_0$ and $d_0$ of the initial condition on the first element in the solution basis. The first admissible solution for $t^*$ thus writes}  
\begin{equation}
	\frac{t^*}{\tau}= \frac{1}{\tilde{\omega}_0}atan2\left(c_0,-\frac{d_0+\alpha c_0}{\tilde{\omega}_0}\right).
	\label{eq:firstRev}
\end{equation} 
{As implicitly reported in equation~\ref{eq:firstRev}, $\beta$ and $De$ (through $E$) influence $t^*$ setting the value of $\alpha$ inside the $atan2$ argument. 
The dependence is non-trivial but, as observed in figure~\ref{fig:4}$(e)$, $t^*$ increases for growing $\beta$ at fixed $E$, and decreases for growing $E$ at fixed $\beta$, within the oscillatory region.}
While we find order-of-magnitude agreement between prediction and simulations, the prediction is systematically around $50\%$ smaller, clearly suggesting that contributions from higher modes are not negligible during the initial transient. {Indeed, in cases where these are absent, the prediction becomes exact (as in the case of a monochromatic excitation on the gravest mode, reported in the Supplementary Information). Furthermore, close to the boundary of the oscillatory region, there are cases with a single reversal followed by monotonic decay (additional details in the Supplementary Information). The least decaying wave assumption is clearly not suitable to capture them as they are not a periodic phenomenon, and likely emerge due to non-linear interactions during the initial transient. Equation~\ref{eq:firstRev} shall thus be considered a mere estimate of the time of the first reversal.}

\subsection{Elastic turbulence}

So far we have focused on the mean--flow in the mixing layer, for which the turbulent fluctuations are seen to be inessential as the simplified model closely reproduces the overturning behaviour. Now, to characterise the fluctuations about the mean--flow, we look at the turbulent kinetic energy $E_t = \boldsymbol{u_t} \cdot \boldsymbol{u_t}/2 $, where $\boldsymbol{u_t}  = \boldsymbol{u}  - \langle \boldsymbol{u}  \rangle_x$; $E_t(\kappa_x)$ is the associated turbulent energy density in the streamwise wavenumber $(\kappa_x)$. 
As shown in the Supplementary Information, $E_t(\kappa_x)$ at the centreline of the mixing layer ($y= 0$) initially has delta function peaks corresponding to the wavenumbers from the initial perturbation, and no scaling for other wavenumbers. Over time a power law scaling develops, with $E_t(\kappa_x) \sim \kappa_x^{-4}$, with the large scale structures associated with the initial condition decaying. This power law scaling has recently been shown to be a signature of elastic turbulence \cite{singh2024intermittency, lellep2024purely, rota2024unified,foggirota-etal-2026}, which is a spatio-temporally chaotic state driven by the balance between the elastic stress from the microstructure and viscous dissipation \cite{singh2024intermittency}.

\section{Discussion}

We have shown that a mixing layer in a viscoelastic fluid exhibits yo-yoing dynamics. Instead of the expected monotonic decay, the entire mixing layer is periodically overturned due to a strong two-way coupling between the {polymers} and the mean--flow. A mono--directional flow model proves capable {of capturing} the yo-yoing dynamics admitting oscillatory solutions in the form of waves, for which we derive an analytical dispersion relation. We thus clarify the role of elasticity and viscosity in the propagation of the observed waves {within} the transient system studied. The yo-yoing vanishes in the zero and, noticeably, also in the infinite elasticity limits.

{The cases reported in our study are dominated by the gravest mode dynamics as (i) the chosen initial condition strongly projects onto it, and (ii) the boundary conditions at the top and bottom of the domain excite all modes, including the gravest one, which becomes dominant if activated. Yet, in potentially unbounded setups, or cases where the boundary conditions preserve modal decoupling (e.g., the periodic case), initial conditions might be implemented so as to isolate the dynamics of higher modes only. We corroborate this statement introducing two additional sets of simulations, initialised from sinusoidal velocity profiles, in the present and a periodic setup. Further details are found in the Supplementary Information.}

{Some limitations of the present study should further be acknowledged. First, the flow reversals reported here occur after a substantial decay of the initial mixing layer, resulting in relatively small residual velocities at the overturning. This behaviour is consistent with the decaying nature of the flow considered and with the low velocity fluctuations typically observed in elastic turbulence \cite{foggirota-etal-2026}. However, our results at lower values of $\beta$ already show that increasing the polymer concentration significantly slows the decay and allows reversals of a more energetic flow (figure~\ref{fig:4}b). We therefore expect larger-amplitude yo-yoing to arise in sustained configurations where energy is continuously supplied to the system. Second, although the analytical predictions depend on the domain size through the selected modal structure, the box-size dependence itself is not intrinsic to the mechanism, but emerges from the combined effect of the chosen initial and boundary conditions. Third, our analysis employs the Oldroyd--B constitutive model and is restricted to two-dimensional simulations. In the present cases however, the polymer extension remains moderate (figure~\ref{fig:1}c), suggesting that finite-extensibility effects are unlikely to play a major role. Likewise, the excellent agreement between the one-dimensional model and the simulated two-dimensional system indicates that the yo-yoing is fundamentally a one-dimensional phenomenon. Nevertheless, assessing the robustness of the mechanism in more realistic constitutive models and fully three-dimensional configurations remains an important direction for future work.}

Furthermore, since mixing layers are the simplest prototype of free shear flows, we expect the striking yo-yoing behaviour to be present in {diverse} unsteady viscoelastic flows. For example, viscoelastic canopy flows are ubiquitous in the biological context \cite{gilpin2020multiscale, hosoi2019corrsin}, yet remain relatively unexplored \cite{de2023canopy, lopez2025canopy}.
Lopez et al. \cite{lopez2025canopy} recently observed the reversal of the shear layer formed by the velocity fluctuations in the elastic flow over micro--canopies, described as a large scale roller with an unknown origin. Our waves provide a justification for those rollers, although future work is required to fully establish the connection. Waves attributed to fluid elasticity have already been reported in channel flows \cite{varshney2018mixing, varshney2019elastic, jha2021elastically}. However, the dispersion relation emerging in our case differs from that reported before \cite{varshney2019elastic} (being nonlinear instead of linear and vanishing for infinite elasticity), ultimately exposing the shear and transient nature of our waves.

We expect our results to be generalisable to mixing layer flows of a range of complex fluids where the microstructure is coupled to the mean shear; some prominent examples being wormlike micelle solutions \cite{cates2006rheology, fardin2010elastic}, suspensions of anisotropic particles \cite{butler2018microstructural, guazzelli2011physical}, liquid crystals \cite{rey2002dynamical}, and active fluids like swimmer suspensions \cite{koch2011collective, guasto2012fluid, lauga2016bacterial}. Finally, the overturning of the two streams can be exploited in microfluidic applications, where mixing enhancement is essential \cite{ottino2004introduction, browne2024harnessing, sasmal2025potential}.

\bibliographystyle{apsrev4-2}
\providecommand{\noopsort}[1]{}\providecommand{\singleletter}[1]{#1}%

\section*{Methods}
The numerical simulations are performed using the in-house solver \emph{Fujin} \cite{rosti-2026}, which has been extensively validated including for viscoelastic fluids \cite{rosti2023large, abdelgawad2023scaling,  singh2024intermittency} (also see \url{https://www.oist.jp/research/research-units/cffu/fujin}). The code uses a staggered, uniform Cartesian grid with a second-order central finite-difference scheme for spatial derivatives and a second order Adams–Bashforth for time integration, coupled with the fractional step method. 
We tackle the evolutive equation for the {polymer conformation tensor} directly, circumventing the high--$De$ instability with a WENO scheme \cite{sugiyama2011full, rosti2023large} for the upper--convected derivative. {Consequently, no polymeric stress diffusion is needed to regularise the problem}. 

At every time step, we find a pressure field satisfying the divergence-free constraint with an efficient spectral Poisson solver, while the overall parallelisation of the code relies on the \emph{2decomp} library along with the message passing interface (MPI) protocol. The two-dimensional domain ($L \delta_0 \times L \delta_0$) is taken to be $L = 28 $, where $\delta_0$ is the width of the mixing layer at $t=0$ discretized with a $256 \times 256 $ grid. The numerical parameters were chosen after verifying convergence with respect to both the grid size and time step. 

The flow is initialised, at $t= 0$, with a hyperbolic tangent mean velocity profile plus a small perturbation \cite{zayernouri2011coherent}, given by,
\begin{align}
\boldsymbol{u} &=  \left( U_0 \tanh \left( 2 y/\delta_0  \right)+ c \frac{\partial \psi}{\partial y} \right) \boldsymbol{\hat{x}},
\label{vel1}
\\
\boldsymbol{v} &=   - c \frac{\partial \psi}{\partial x} \boldsymbol{\hat{y}},
\label{vel2}
\end{align}
where $c=0.001$ and $\psi(x,y) = \mathrm{exp}(-y^2/\delta_0^2)[\cos(8\pi x) + \cos(20 \pi x)]$. Initially, the mixing layer has two streams which are moving in opposite directions with speed $U_0$ and thickness $\delta_0$. In cases where ensemble averaging is required, a random number, with a small magnitude, is added to the initial condition. We have verified that such noise has no affect on the mean dynamics.

\section*{Data availability statement}
All data needed to evaluate the conclusions are present in the manuscript and/or the Supplementary Information.
Data required to reproduce the figures are available on the website of the Complex Fluids and Flows Unit at OIST (\url{https://www.oist.jp/research/research-units/cffu/publications/publication-data}), where full details about the code used for the present research are also found. 

\section*{Author contributions}
MER conceived the original idea, refined upon discussion with GFR and PG. JT performed the numerical simulations and preliminary analysed the data, supervised by GFR and PG. GFR, PG, and MER derived the simplified model, solved and analysed by GFR. PG wrote the initial draft, with input from GFR and MER; GFR finalised the manuscript. MER supervised the project.

\begin{acknowledgments}
The research was supported by the Okinawa Institute of Science and Technology Graduate University (OIST) with subsidy funding to MER from the Cabinet Office, Government of Japan. MER also acknowledges funding from the Japan Society for the Promotion of Science (JSPS), grants 24K00810 and 24K17210. The authors acknowledge the computer time provided by the Scientific Computing \& Data Analysis section of the Core Facilities at OIST, and GFR and MER also acknowledge the HPCI Research Project grants hp250035, hp260009, and hp260019.\\

GFR, JT, and MER acknowledge with sadness the passing of PG during the course of this work, and dedicate this paper to his memory.
\end{acknowledgments}

\end{document}



\setcounter{table}{0}
\makeatletter 
\renewcommand{\thetable}{S\@arabic\c@table}
\makeatother

\setcounter{figure}{0}
\makeatletter 
\renewcommand{\thefigure}{S\@arabic\c@figure}
\makeatother

\setcounter{equation}{0}
\makeatletter 
\renewcommand{\theequation}{E\@arabic\c@equation}
\makeatother


\title{\small Supplementary information for the paper \\ \vspace{.5cm} \large Waves dictate the yo-yoing decay of a viscoelastic mixing layer}


\author{Giulio Foggi Rota\orcidB{}}
\thanks{These authors contributed equally to this work.}
\affiliation{Complex Fluids and Flows Unit, Okinawa Institute of Science and Technology Graduate University, Okinawa 904-0495, Japan}
\author{Piyush Garg\orcidC{}}
\thanks{These authors contributed equally to this work.}
\affiliation{Complex Fluids and Flows Unit, Okinawa Institute of Science and Technology Graduate University, Okinawa 904-0495, Japan}
\author{Jason Tang\orcidA{}}
\affiliation{Complex Fluids and Flows Unit, Okinawa Institute of Science and Technology Graduate University, Okinawa 904-0495, Japan}
\author{Marco Edoardo Rosti\orcidD{}}%
\email{marco.rosti@oist.jp}
\affiliation{Complex Fluids and Flows Unit, Okinawa Institute of Science and Technology Graduate University, Okinawa 904-0495, Japan}

\maketitle

\section{Derivation of the energy balance}
\begin{figure}
		\includegraphics[width=.5\textwidth]{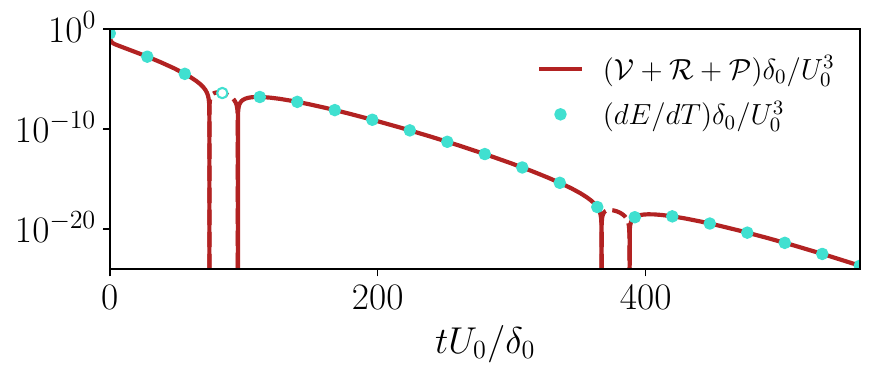}
	\caption{Various terms of the energy balance equation \eqref{eq:MKE} over time $tU_0/\delta_0$; where $Re=0.2$, $De = 28$, $\beta = 0.9$.} 
    \label{fig:energy balance}
\end{figure}

Define $\boldsymbol{S} =  \frac{1}{2}(\nabla\mathbf{u} + (\nabla\mathbf{u})^T)$, and the turbulent velocity as $\boldsymbol{u}_t =\boldsymbol{u} - \langle \boldsymbol{u}\rangle_x $.
Taking the $x$-average of the momentum equation gives
\begin{equation}
\partial_t \langle \mathbf{u} \rangle_x  = \nabla \cdot \left(- \frac{\langle p \rangle_x}{\rho_f} \mathbf{I} + \frac{\mu_p}{\rho_f\tau_p} \langle \boldsymbol{C} \rangle_x  +2 \frac{\mu_f}{\rho_f} \langle \mathbf{S} \rangle_x - \langle \mathbf{u}_t \mathbf{u}_t \rangle_x -  \langle \mathbf{u} \rangle_x \langle \mathbf{u} \rangle_x \right),
\end{equation} 
and taking a dot product with $\langle \mathbf{u} \rangle_x$ and rearranging we obtain
\begin{equation}
\partial_t \left(\frac{1}{2}\langle \mathbf{u} \rangle_x\cdot\langle \mathbf{u} \rangle_x \right)   = - \left(\frac{\mu_p}{\rho\tau_p} \langle \boldsymbol{C} \rangle_x  +2 \frac{\mu_f}{\rho}  \langle \mathbf{S} \rangle_x - \langle \mathbf{u}_t \mathbf{u}_t \rangle_x \right)  \boldsymbol{:}  \langle \nabla \mathbf{u} \rangle_x  + \nabla \cdot ( \text{flux}). 
\end{equation} 
Integrating over the whole domain and assuming no flux of kinetic energy across the boundaries yields the energy balance equation as
\begin{equation}
  \frac{\mathrm{d}E}{\mathrm{d}t} = -  \biggl \langle \left( 2 \frac{\mu_f}{\rho} \langle \mathbf{S} \rangle_x  -   \langle {\mathbf{u}_t \mathbf{u}_t} \rangle_x + \frac{\mu_p}{\rho \tau_p} \langle \mathbf{C} \rangle_x \right) \boldsymbol{:} \langle \nabla  \mathbf{u} \rangle_x  \biggr\rangle,
  \label{eq:MKE}
\end{equation}
which is stated in the main text. Figure \ref{fig:energy balance} verifies that the energy balance in \eqref{eq:MKE} is satisfied exactly for all times $t$ for which results are presented, thus demonstrating the accuracy of the numerical scheme. 

\section{Negligible role of the advective non-linearity}


To demonstrate that the advective non-linearity plays a negligible role in the yo-yoing dynamics, we carried out a simulation where it is set to exactly zero in the momentum equation, but the time derivative term is not. Figure \ref{fig:no perturbation} compares the time evolution of the mean {centreline} vorticity {$| \langle \omega \rangle_c  / \langle \omega_0 \rangle_c |$} when the advective term ($\boldsymbol{u} \cdot \nabla \boldsymbol{u}$) is included and excluded from the simulation. The results are seen to be essentially unchanged.

\section{Negligible role of the initial perturbation}

\begin{figure}
		\includegraphics[width=.475\textwidth]{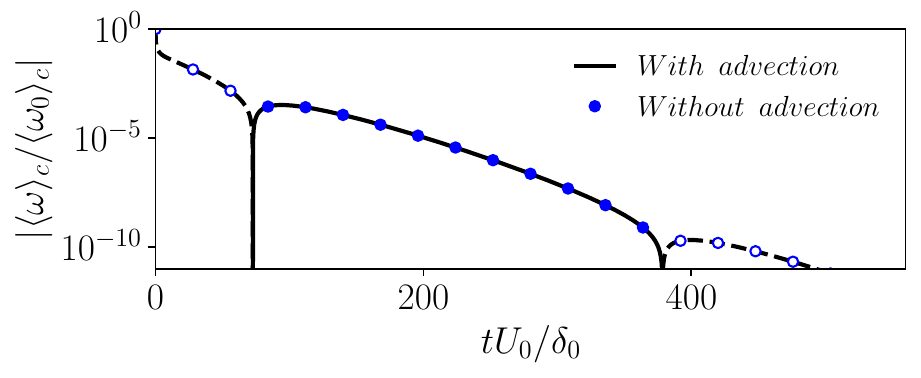}
		\includegraphics[width=.475\textwidth]{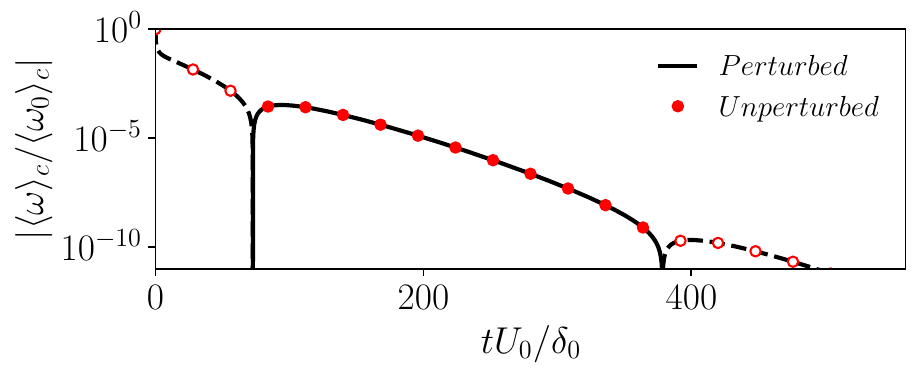}
	\caption{ \underline{Left}: evolution of the {mean centreline} vorticity ({$| \langle \omega \rangle_c  / \langle \omega_0 \rangle_c |$}) over time ($tU_{0}/\delta_0$) for the mixing layer flow with and without advective non-linearity; $Re=0.2$, $De = 28$, $\beta = 0.9$. \underline{Right}: same for the two cases with and without an additional perturbation to the mixing layer at $t=0$; same parameters.}
    \label{fig:no perturbation}
\end{figure}

Next, we verify that the initial perturbation to the mixing layer plays a negligible role in the yo-yoing dynamics by removing it. Figure \ref{fig:no perturbation} compares the mean {centreline} vorticity ${| \langle \omega \rangle_c  / \langle \omega_0 \rangle_c |}$ for the cases with and without an initial perturbation to the mixing layer. We can see that the overturning of the mixing layer is not dependent on the initial perturbation.

\begin{figure}[b]
		\includegraphics[width=.5\textwidth]{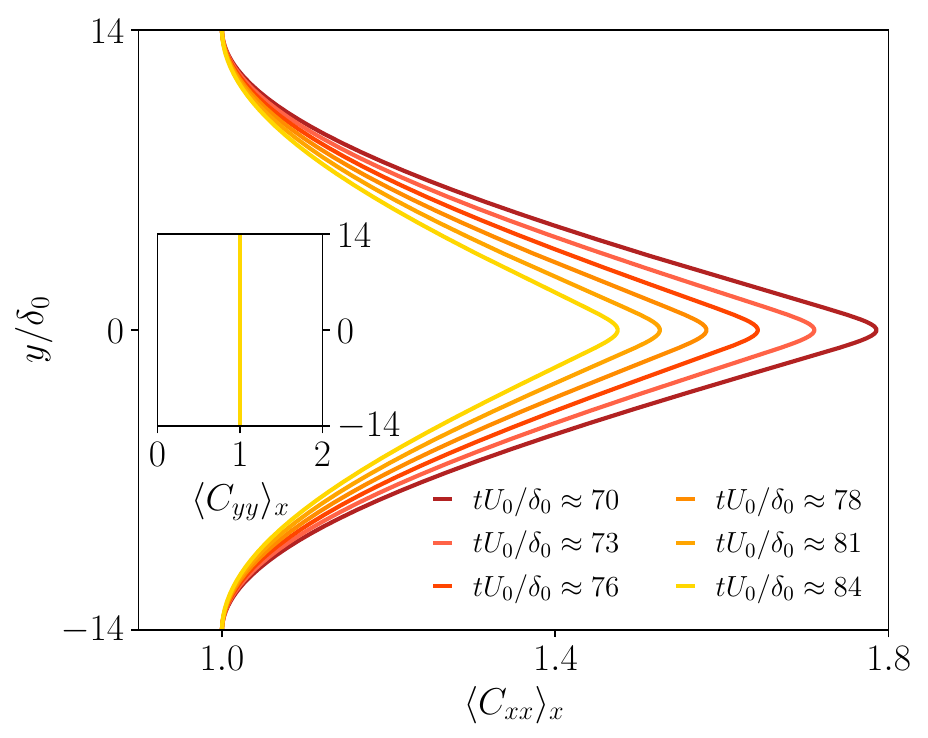}
	\caption{The spatial variation of $\langle C_{xx} \rangle_x$ versus $y/\delta_0$ at various times $tU_0/\delta_0$ near the overturning time; the inset shows the unchanging $\langle C_{yy} \rangle_x$. The parameters are the same as the main paper, i.e. $Re=0.2$, $De = 28$, $\beta = 0.9$.}  
    \label{fig:Cxx}
\end{figure}

\section{Reduced model for the flow and the polymer field}
\begin{figure}
		\includegraphics[width=0.475\textwidth]{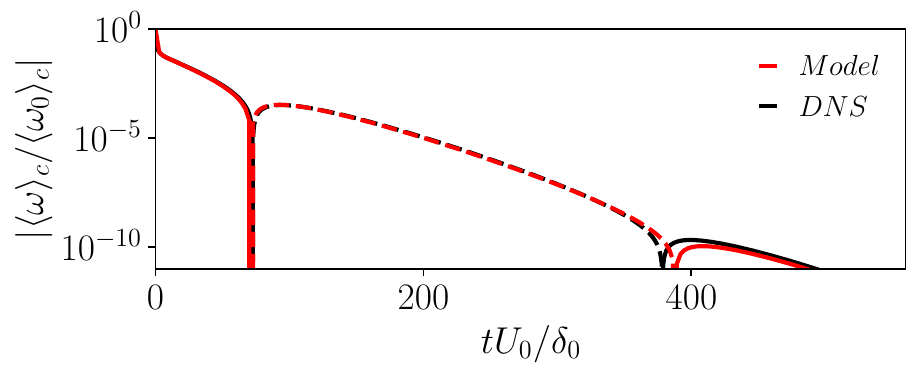}
		\includegraphics[width=0.475\textwidth]{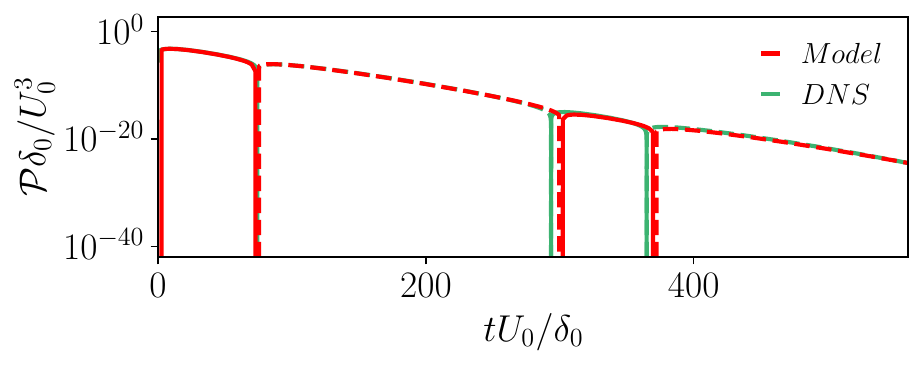}
	\caption{\underline{Left}: validation of the reduced 1D-model for the time evolution of the {mean centreline} vorticity against numerical simulation results; $Re=0.2$, $De = 28$, $\beta = 0.9$.
	\underline{Right}: validation of the reduced 1D-model for the time evolution of the polymer production term in the energy balance against numerical simulation results, same parameters.}
    \label{fig:valid_1}
\end{figure}

In the main text, we have shown the streamwise-averaged $\langle u \rangle_x$ and $\langle C_{xy} \rangle_x$. Figure \ref{fig:Cxx} shows the streamwise average of the other two components of the {conformation} tensor ($\langle C_{xx} \rangle_x$ and $\langle C_{yy} \rangle_x$). Taken together, the simulation results suggest that the turbulent fluctuations do not affect the streamwise-averaged flow and {polymer field}. On the assumption of unidirectional streamwise averaged flow, $\langle v \rangle_x = 0$ and $\langle u \rangle_x$ is only a function of the gradient coordinate, $y$. We also assume that the streamwise averaged {conformation tensor} $\langle \boldsymbol{C} \rangle_x$ is only a function of $y$, which we have also verified from our simulation results. The governing equations for the streamwise averaged flow field thus simplify to
\begin{align}
    \langle v \rangle_x  &= 0\\
    \frac{\partial\langle u \rangle_x}{\partial t} &= \frac{\mu_f}{\rho} \frac{\partial^2\langle u \rangle_x}{\partial y^2} +  \frac{\mu_p}{\rho \tau_p}\frac{\partial \langle C_{xy} \rangle_x}{\partial y}, \label{eq:u} \\  
    \frac{\partial \langle C_{xy} \rangle_x}{\partial t} &= \langle C_{yy} \rangle_x \frac{\partial\langle u \rangle_x}{\partial y} - \frac{1}{\tau_p}\langle C_{xy} \rangle_x, \label{eq:Cxy} \\
    \frac{\partial \langle C_{xx} \rangle_x}{\partial t} &= \langle C_{xy} \rangle_x \frac{\partial\langle u \rangle_x}{\partial y} - \frac{1}{\tau_p}(\langle C_{xx} \rangle_x -1), \label{eq:Cxx} \\
    \frac{\partial \langle C_{yy} \rangle_x}{\partial t}  &= -\frac{1}{\tau_p}(\langle C_{yy} \rangle_x -1).  \label{eq:Cyy}
\end{align}
\eqref{eq:Cyy} is trivially solved with $C_{yy} = 1$. $\langle u \rangle_x$ and $\langle C_{xy} \rangle_x$ evolve through the coupled equations \eqref{eq:u} and  \eqref{eq:Cxy}, and $\langle C_{xx}\rangle_x$ can then be obtained from \eqref{eq:Cxx}. We can also derive a reduced equation for $\langle u \rangle_x$ as,
\begin{equation}
   \frac{\partial^2 \langle u\rangle_x }{\partial t^2} +  \frac{1}{\tau_p} \frac{\partial \langle u_{} \rangle_x}{\partial t}   - \frac{\mu_f + \mu_p}{\rho \tau_p}  \frac{\partial^2 \langle u\rangle_x }{\partial y^2} = \frac{\mu_f}{\rho} \frac{\partial^3 \langle u\rangle_x }{\partial t \partial y^2 }.
   \label{eq:red}
\end{equation}
Equation \eqref{eq:red} has been derived earlier in the context of Stokes' first problem for viscoelastic fluids \cite{christov2010stokes}, but here it is instead subject to the initial and boundary conditions corresponding to the mixing layer.
To verify if the reduced set of equations is a good model for the overturning, we numerically integrate \eqref{eq:u} and  \eqref{eq:Cxy} in time. The two approaches - DNSs and 1D-model - show good agreement, see figure \ref{fig:valid_1}, thus the overturning of the mixing layer is primarily a mean flow phenomenon. 

After rewriting \eqref{eq:red} in dimensionless form as reported in the main text, 
\begin{align}
   \frac{\partial^2 \tilde{u} }{\partial \tilde{t}^2} +  \frac{\partial \tilde{u}}{\partial \tilde{t}}   - E  \frac{\partial^2 \tilde{u} }{\partial \tilde{y}^2} = E \beta \frac{\partial^3 \tilde{u} }{\partial \tilde{t} \partial \tilde{y}^2 },
   \label{eq:redAdim}
\end{align}
(where $\tilde{y}=y/\delta_0$, $\tilde{t}=t /\tau$, $\tilde{u}=u/U_0$, and $E=De/Re$ is the elasticity number), we proceed to solve it analytically for $\tilde{y}\in[-L/2;L/2]$ and $\tilde{t}\in(0,+\infty)$. 
We require $\tilde{u}(y,0)=\tanh(2\tilde{y})$ and $\lim_{\tilde{t}\to\infty} u(\tilde{y},\tilde{t})=0$, while $\displaystyle \frac{\partial \tilde{u}}{\partial \tilde{y}}\bigg|_{\tilde{y}=\pm L/2}=0$. 

In the limit $E\rightarrow 0$, \eqref{eq:redAdim} reduces to 
\begin{equation}
     \frac{\partial}{\partial \tilde{t}}\left( \frac{\partial\tilde{u}}{\partial \tilde{t}} + \tilde{u} \right)=0,
\end{equation}
which admits a monotonously decaying solution $\tilde{u} = \tanh(2\tilde{y})e^{-\tilde{t}}$; however note that this only satisfies the boundary condition for special initial conditions, thus implying the singular nature of this limit. 

In the limit $E\rightarrow +\infty$, \eqref{eq:redAdim} reduces to 
\begin{equation}
    -\displaystyle \frac{\partial^2\tilde{u}}{\partial \tilde{y}^2}=\beta \frac{\partial^3\tilde{u}}{\partial \tilde{t} \partial^2 \tilde{y}},
\end{equation}
which also admits a monotonously decaying solution $\tilde{u} = \tanh(2\tilde{y})e^{-\tilde{t}/\beta}$ (assuming that the initial condition satisfies the boundary conditions). 

More interesting is the case of the full equation \eqref{eq:redAdim}, which we tackle separating $\tilde{u}(\tilde{y},\tilde{t})=A(\tilde{t})B(\tilde{y})$. We thus rewrite 
\begin{align}
   (E \beta A' + E A) B'' = (A'+A'')B,
   \label{eq:vSep}
\end{align}
where primes denote differentiation relative to the relevant variable. For $B\neq0$ and $E \beta A' + E A \neq 0$, introducing a properly chosen set of wavenumbers $\tilde{k}_n \in \mathbb{R}_+$ with $n \in \mathbb{N}_0$ on which an arbitrary initial condition projects, we split \eqref{eq:vSep} into
\begin{align}
   \frac{A'+A''}{E \beta A' + E A} = -\tilde{k}_n^2, \enspace \frac{B''}{B}=-\tilde{k}_n^2.
\end{align}
The equation for $A(\tilde{t})$ rearranges as $A''+(1+\beta E \tilde{k}_n^2) A' + E \tilde{k}_n^2 A =0$, with characteristic roots
\begin{align}
   \gamma_\pm (\tilde{k}_n) = - \displaystyle \frac{1+\beta E \tilde{k}_n^2}{2} \pm \sqrt{\left( \frac{1+\beta E \tilde{k}_n^2}{2} \right)^2-E \tilde{k}_n^2}
   \label{eq:exps}
\end{align} 
governing the time evolution of the mixing layer. In particular, oscillations of the $n^{th}$ mode occur if $\Im(\gamma_\pm (\tilde{k}_n))\neq0$, with frequency 
\begin{align}
   \tilde{\omega}_n = \sqrt{E \tilde{k}_n^2 - \left(\frac{1+\beta E \tilde{k}_n^2}{2}\right)^2},
   \label{eq:disp}
\end{align}
as reported in the main text. Noting that the periodic overturning of the mixing layer we observed is mostly a $\tilde{k}_0$ dynamics, we estimate the oscillation period as $T/\tau=2\pi/\tilde{\omega}_0$. \\
The equation for $B(\tilde{t})$ admits sinusoidal solutions $B_n(\tilde{t})$ for each mode $\tilde{k}_n$, so that the general solution to \eqref{eq:redAdim} writes
\begin{align}
   \tilde{u}(\tilde{y}, \tilde{t})=\sum_{n=0}^{+\infty} \left(a_n e^{\gamma_+(\tilde{k}_n)\tilde{t}} + b_n e^{\gamma_-(\tilde{k}_n)\tilde{t}}\right) B_n(\tilde{y}),
   \label{eq:sol}
\end{align}
with the coefficients $a_n, b_n \in \mathbb{R}$ fixed by the initial and boundary conditions.

\begin{figure}
	\includegraphics[width=0.5\textwidth]{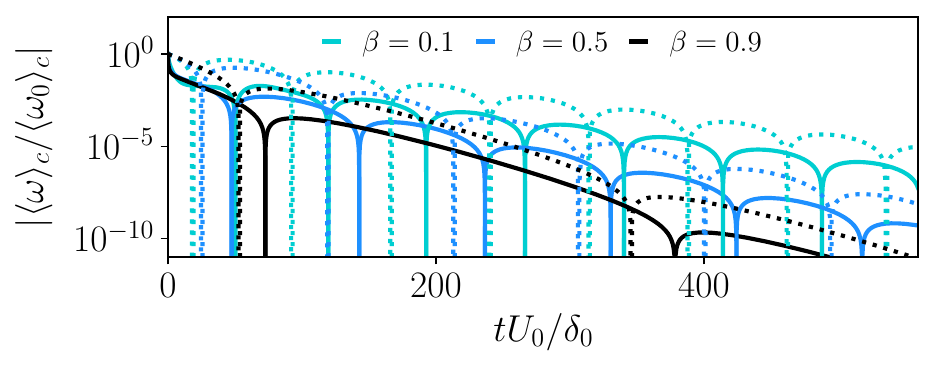}
	\caption{{Comparison between the first active mode dynamics predicted by the analytical solution (dotted lines) and numerical simulation results (continuous lines). We show the mean centreline vorticity for the case at $Re=0.2$ and $De = 28$. The period and the decay rate are matched exactly, with the only discrepancy being the time of the first reversal as a consequence of the initial transient.}}
	\label{fig:analytical}
\end{figure}

For completeness, we now specialise the general solution \eqref{eq:sol} to our case. As the initial condition is an odd function, we select basis functions of the form 
\begin{align}
   B_n(\tilde{y}) = \sin\left(\frac{(2n+1)\pi\tilde{y}}{L}\right) \text{ with wavenumbers } \tilde{k}_n=\frac{(2n+1)\pi}{L},
\end{align}
compatible with the boundary conditions.
{At $\tilde{t}=0$ we require $\tilde{u} (\tilde{y},0) = \tilde{u}_0 =  tanh(2\tilde{y})$ and impose the diffusive balance $\displaystyle\frac{\partial\tilde{u}}{\partial \tilde{t}}(\tilde{y},0)= E \displaystyle\frac{\partial^2 \tilde{u}_0}{\partial \tilde{y}^2}$, so the normalised projections of the initial conditions on the elements of the base write
\begin{align}
   c_n = \frac{\int_{-L/2}^{L/2} \tanh(2\tilde{y}) B_n(\tilde{y}) d\tilde{y}}{\int_{-L/2}^{L/2} B_n^2(\tilde{y}) d\tilde{y}}, \enspace \textrm{and} \enspace d_n = \frac{\displaystyle\int_{-L/2}^{L/2} E \displaystyle\frac{\partial^2  \tanh(2\tilde{y})}{\partial \tilde{y}^2}  B_n(\tilde{y}) d\tilde{y}}{\int_{-L/2}^{L/2} B_n^2(\tilde{y}) d\tilde{y}}.
\end{align}
We may finally isolate the coefficients $a_n, b_n$ as
\begin{align}
   a_n=\frac{d_n-c_n\gamma_-(\tilde{k}_n)}{\gamma_+(\tilde{k}_n)-\gamma_-(\tilde{k}_n)} \text{ and } b_n=\frac{c_n\gamma_+(\tilde{k}_n)-d_n}{\gamma_+(\tilde{k}_n)-\gamma_-(\tilde{k}_n)}.
\end{align}
}

{To corroborate the understanding that the periodic overturning of the shear layer is dictated by the gravest active mode, in figure~\ref{fig:analytical} we plot the analytical solution just found truncated at $n=0$ as dotted lines, versus the outcome of our simulations. After the initial adaptation of the simulations, which entails complex modal dynamics not captured by the \textit{least decaying wave assumption}, both the period of the reversals and the decay rate are matched exactly. The time of the first reversal predicted by the model, instead, might be taken as an order-of-magnitude estimate of the simulated one, but systematically under-predicts it by around $50\%$, suggesting that contributions from higher modes are non-negligible during the initial transient.
}




{\section{Occurrence of cases with a single reversal}}

\begin{figure}
   \includegraphics[width=0.5\textwidth]{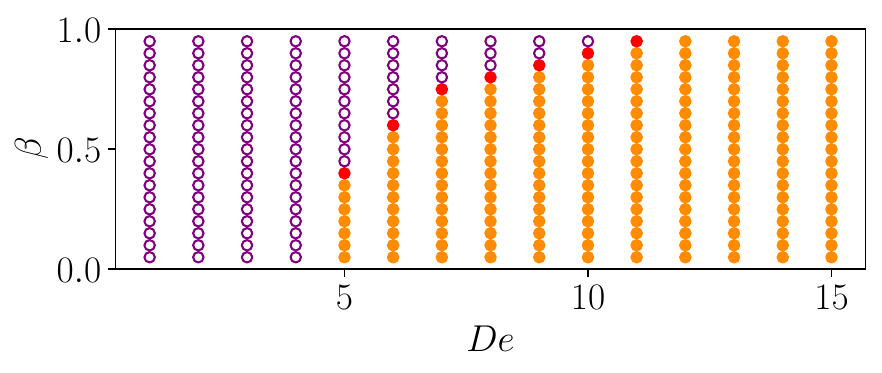}
   \caption{{Subset of the $De-\beta$ space for $Re=0.2$ highlighting cases with a single flow reversal followed by monotonic decay (red dots) at the boundary between oscillatory (orange, full) and non-oscillatory (violet, empty) cases.}}
   \label{fig:1osc}
\end{figure}

{Cases with a single reversal followed by monotonic decay systematically occur close to the boundary of the oscillatory region in the parameter space. 
The least decaying wave assumption is not suitable to capture them, so we instead solve our flow model spanning a subset of the $De-\beta$ space for a chosen value of $Re$, and report the outcome in figure~\ref{fig:1osc}. It clearly appears that such cases are indeed located at the aforementioned boundary, and likely emerge due to non-linear interactions during the initial transient.} \\

{\section{Effects of the initial and boundary conditions}}

\begin{figure}
   \includegraphics[width=.46\textwidth]{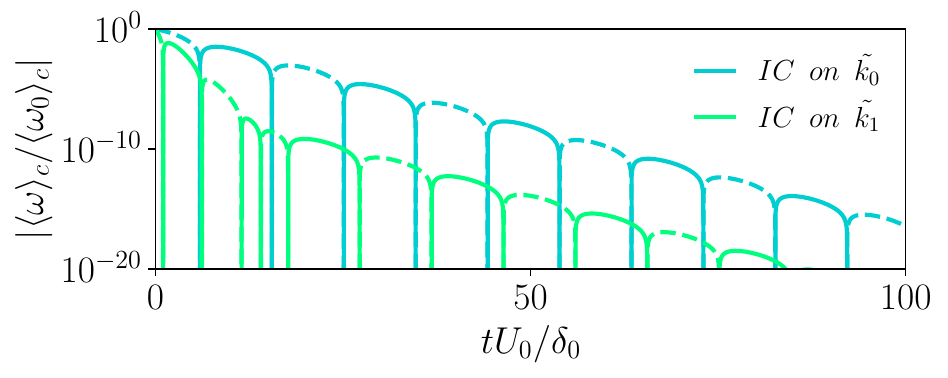}
   \includegraphics[width=.44\textwidth]{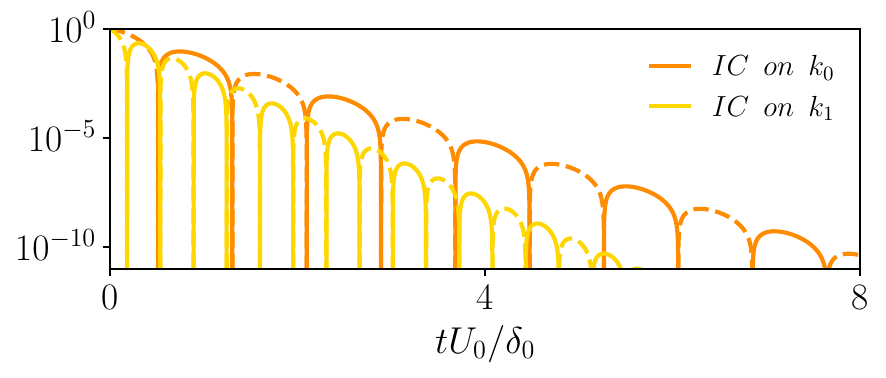}
   \caption{{\underline{Left}: evolution of sinusoidal initial conditions (ICs) imposed on $\tilde{k}_0$ (blue) and $\tilde{k}_1$ (green) in the setup of our study. \underline{Right}: evolution of sinusoidal initial conditions imposed on ${k}_0$ (orange) and ${k}_1$ (gold) in a mono-directional Kolmogorov's flow. Continuous lines denote positive values of $\omega$, dashed lines negative ones.}}
   \label{fig:tests}
\end{figure}

{First we consider a box with edge $L=2\delta_0$, periodic along the streamwise direction, bounded along the gradient direction by no-penetration and free-slip conditions on the velocity, along with null second derivative of the conformation tensor (as in the rest of our study). We select parameters in the oscillatory regime, characterised by $Re=10$, $De=2$, and $\beta=0.5$, and evolve the flow starting from two distinct initial conditions (ICs): a sinusoid of the form $u= U_0 \sin(\tilde{k}_0 y)$ only triggering the gravest mode, and a sinusoid of the form $u= U_0 \sin(\tilde{k}_1 y)$ only triggering the mode immediately above. The trend of the centreline vorticity is reported in the left of figure~\ref{fig:tests}. As expected, when the IC is provided on $\tilde{k}_0$, regular oscillations develop according to our theoretical predictions ($T U_0/\delta_0 \approx 19.4$ and first reversal at $t^*U_0/\delta_0 \approx 6$); note that in this case also the time of the first reversal is matched as contributions from higher modes do not emerge. Instead, upon providing the IC on $\tilde{k}_1$, a different initial behaviour is observed, likely linked to the flow response at that mode. Yet, the chosen boundary conditions also excite dynamics on $\tilde{k}_0$ with a slower decay, that eventually emerge and dominate the flow response. This demonstrates that initial conditions might be purposely designed to excite specific modes, but eventually the chosen setup triggers space-filling dynamics and those become dominant over finite times.} 
 
{Next, to demonstrate the boundary conditions (BCs) effect, we shift our attention to a {mono-directional Kolmogorov's flow}, considering a box with edge $L=4\delta_0$ periodic along the streamwise and gradient directions. Once again, we select parameters in the oscillatory regime, characterised by $Re=2.5$, $De=0.5$, and $\beta=0.2$, and evolve the flow starting from two distinct initial conditions (ICs): a sinusoid of the form $u= U_0 \sin(k_0 y)$, with $k_0=2\pi/L$, only triggering the gravest mode, and a sinusoid of the form $u= U_0 \sin(k_1 y)$, with $k_1=4\pi/L$, only triggering the mode immediately above. The trend of the centreline vorticity is reported in the right of figure~\ref{fig:tests}. In this case, the different initial conditions trigger different modal responses, preserved throughout the decay since the periodic boundaries do not excite additional modal dynamics.} \\
 
\section{Turbulent kinetic energy spectra}
\begin{figure}
		\includegraphics[width=.5\textwidth]{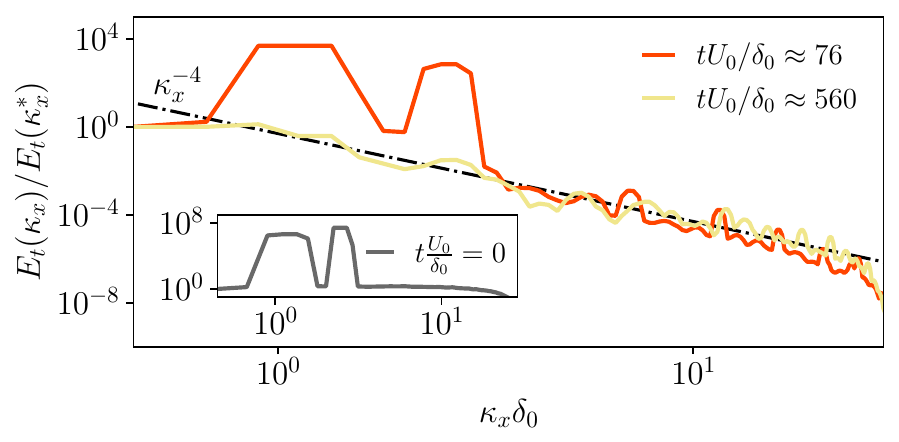}
	\caption{Turbulent kinetic energy density $E_t(\kappa_x)$ versus the non-dimensional streamwise wavenumber $\kappa_x \delta_0$ for two different times $tU_0/\delta_0$. The inset shows the energy spectrum at $t=0$. Note that $E_t(\kappa_x)$ is normalised by the value for the wavenumber $\kappa_x^*=2\pi/(14\delta_0)$ for each $tU_0/\delta_0$.}
    \label{fig:spectra}
\end{figure}

We report in figure \ref{fig:spectra} the spectra of the turbulent kinetic energy $E_t(\kappa_x)$ at the centerline of the mixing layer ($y= 0$) for various times $tU_0/\delta_0$, as discussed in the main text.

\section{Additional results for the polymer stress}
\begin{figure}
		\includegraphics[width=0.8\textwidth]{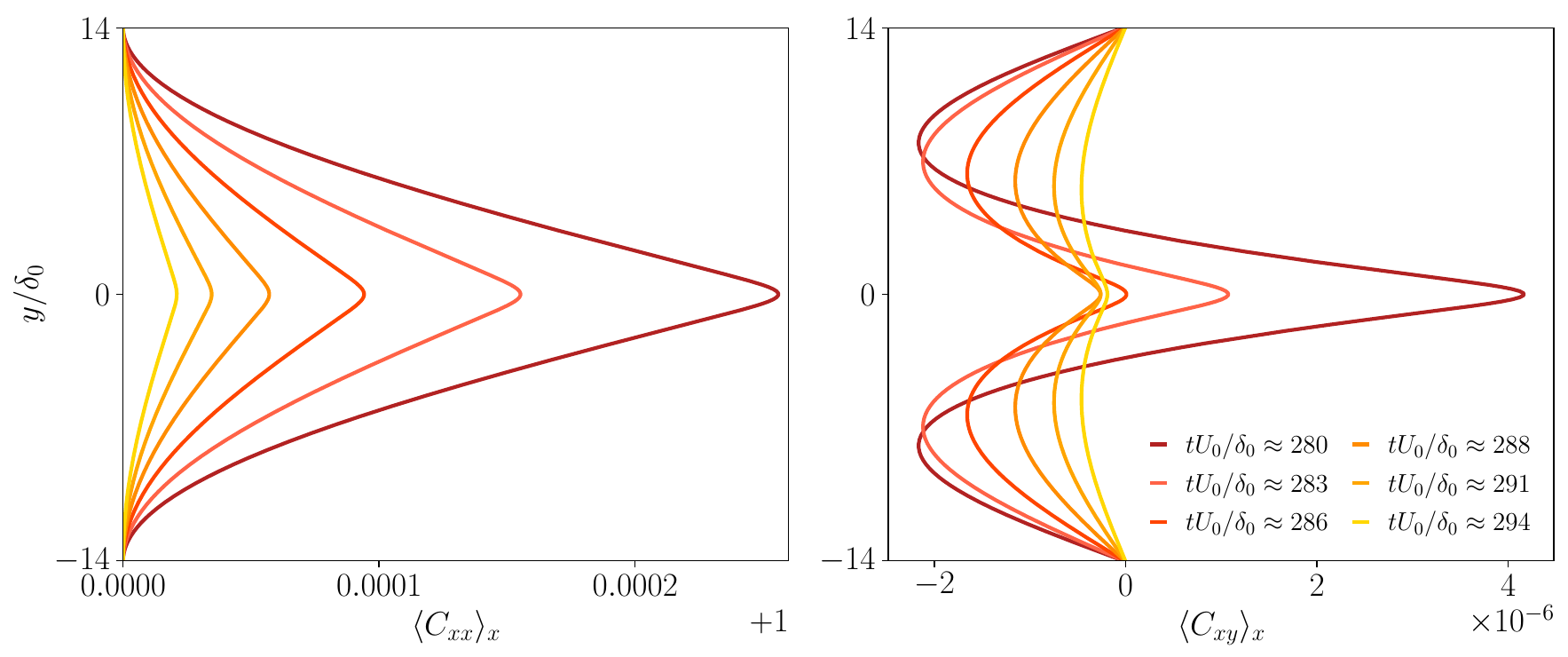}
	\caption{\underline{Left}: spatial variation of  $\langle C_{xx} \rangle_x$ versus $y/\delta_0$ at various times near $t U_0/\delta_0\sim 293.5$; the parameters are the same as the main paper, i.e. $Re=0.2$, $De = 28$, $\beta = 0.9$. \underline{Right}: same for $\langle C_{xy} \rangle_x$. }
    \label{fig:polyOverturn}
\end{figure}

\begin{figure}[h]
		\includegraphics[width=1\textwidth]{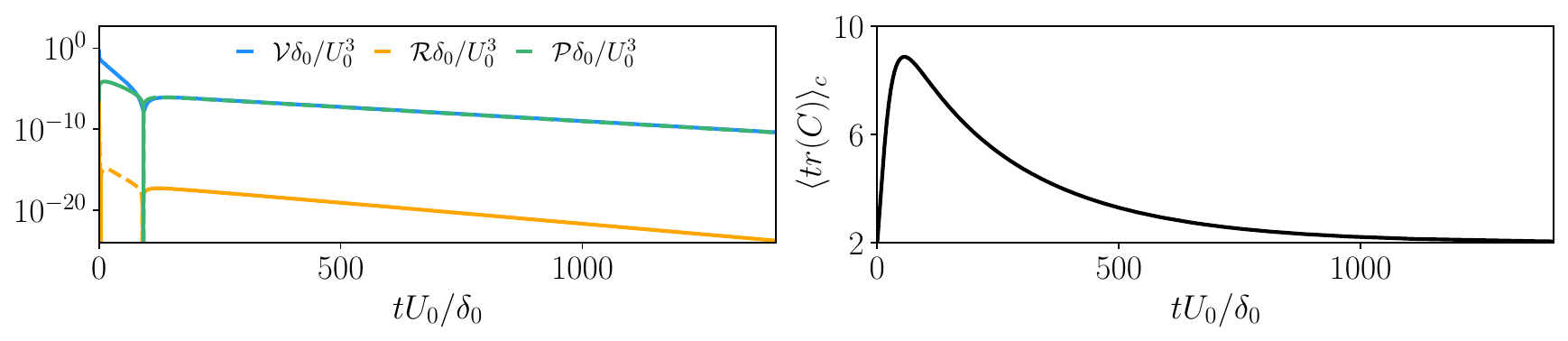}
	\caption{\underline{Left}: time evolution of the various terms in the energy balance. \underline{Right}: time evolution of {$\langle \mathrm{tr}(\mathbf{C})\rangle_c$} for the large $De = 280$ case; $Re=0.2$, $\beta = 0.9$.}
    \label{fig:largeDe}
\end{figure}

In this section, we present some additional results for the {polymer conformation tensor}. Fig \ref{fig:polyOverturn} shows the behaviour of the {polymers} near $t U_0/\delta_0 \sim 293.5$ when the polymer production term again changes sign. From the change in the spatial distribution of $\langle C_{xy} \rangle_x$, we can see that at this time the {polymers} gets reoriented along the flow and hence the polymer production changes sign to be again dissipative. Finally, figure \ref{fig:largeDe} shows the time evolution of the various terms in the energy balance, as well as {$\langle \mathrm{tr}(\mathbf{C})\rangle_c$}, for the large $De$ case. Here, we find that the decay of the mixing layer after the overturning is slower, since the contribution of the polymer production term is larger. 
 
\providecommand{\noopsort}[1]{}\providecommand{\singleletter}[1]{#1}%